\def\re#1{(\ref{#1})}
\def\beq{\begin{equation}}
\def\eeq{\end{equation}}
\def\beeq{\begin{eqnarray}}
\def\beeqn{\begin{eqnarray*}}
\def\eeeq{\end{eqnarray}}
\def\eeeqn{\end{eqnarray*}}

\def\de{\delta}                 \def\D{\Delta}

\def\l{\lambda}                 
\def\m{\mu}
\def\n{\nu}

\newcommand{\DD}{{\cal D}}

\newcommand{\WW}{{\cal W}}

\newcommand{\lp}{\left(}
\newcommand{\rp}{\right)}
\renewcommand{\lq}{\left[}
\renewcommand{\rq}{\right]}

\newcommand{\no}{\nonumber}
\newcommand{\ph}{\phantom} 

\def\tr{\,\mbox{Tr}\,}
\def\frac#1#2{ {{#1} \over {#2} }}

\def\p{\partial}

\newcommand{\unity}{1\kern-.65mm \mbox{\form l}}
\newcommand{\ks}{\mbox{\form l}\kern-.6mm \mbox{\form K}}
\newcommand{\A}{A \raise0.5mm\hbox{\kern-1.8mm /}}
\def\pmb#1{\leavevmode\setbox0=\hbox{$#1$}\kern-.025em\copy0\kern-\wd0
\kern-.05em\copy0\kern-\wd0\kern-.025em\raise.0433em\box0}
\def\D{\hbox{\hbox{${D}$}}\kern-1.9mm{\hbox{${/}$}}}
\def\kbar{\hbox{$k$}\kern-0.2true cm\hbox{$/$}}
\def\nbar{\hbox{$n$}\kern-0.23true cm\hbox{$/$}}
\def\pbar{\hbox{$p$}\kern-0.18true cm\hbox{$/$}}
\def\nhbar{\hbox{$\hat n$}\kern-0.23true cm\hbox{$/$}}

\documentstyle[epsfig,preprint,aps,floats,amssymb]{revtex}
\begin{document}
\draft
\newfont{\form}{cmss10}

\title{Scaling properties of the perturbative Wilson loop in 
two-dimensional non-commutative Yang-Mills theory}
\author{A. Bassetto$^1$, G. Nardelli$^2$ and A. Torrielli$^1$}
\address{$^1$Dipartimento di Fisica ``G.Galilei", Via Marzolo 8, 35131
Padova, Italy\\
INFN, Sezione di Padova, Italy \texttt{(bassetto, torrielli@pd.infn.it)}\\
$^2$Dipartimento di Fisica, Universit\`a di Trento,
38050 Povo (Trento), Italy \\ INFN, Gruppo Collegato di Trento, Italy 
\texttt{(nardelli@science.unitn.it)}}
\maketitle
\begin{abstract}
Commutative Yang-Mills theories in 1+1 dimensions exhibit an interesting
interplay between geometrical properties and $U(N)$ gauge structures: in
the exact expression of a Wilson loop with $n$ windings a non trivial
scaling intertwines $n$ and $N$. In the non-commutative case the interplay
becomes tighter owing to the merging of space-time and ``internal''
symmetries in a larger gauge group $U(\infty)$. We perform an explicit perturbative 
calculation of such a loop up to ${\cal O}(g^6)$; rather surprisingly,
we find that in the contribution from the crossed graphs (the genuine 
non-commutative terms) the scaling we mentioned occurs for large $n$ and 
$N$ in the limit of maximal non-commutativity $\theta=\infty$. 
We present arguments in favour of the persistence of such a scaling
at any perturbative order and succeed in summing the related perturbative series.

\end{abstract}
\vskip 0.5truecm
DFPD 02/TH/12, UTF 448.

\noindent
PACS numbers: 11.15.Bt, 02.40.Gh

\noindent
{\it Keywords}: Non-commutative gauge theories, 
Wilson loops.

\vfill\eject

\narrowtext

\section{Introduction}
\noindent
One of the most interesting and intriguing features of noncommutative
field theories is the merging of space-time and ``internal'' symmetries
in a larger gauge group $U(\infty)$ \cite{alvar,har}. Peculiar topological
properties can find their place there and be conveniently described
under the general frame provided by the K-theory \cite{book}.

On the other hand some interplay occurs also when theories are defined
on commutative spaces; in \cite{BGV} it has been shown that in two space-time
dimensions a non trivial
holonomy concerning the base manifold and the fiber $U(N)$
appears when considering a Wilson loop winding
$n$ times around a closed contour, leading to a peculiar scaling law
intertwining the two integers $n$ and $N$
\begin{equation}
\label{scaling}
\WW_n({\cal A};N)=\WW_N(\frac{n}{N}{\cal A};n),
\end{equation}
$\WW$ being the exact expression of the Wilson loop and
${\cal A}$ the enclosed area.
When going around the loop the non-Abelian character of the
gauge group is felt.

One may wonder whether similar relations are present in the noncommutative
case and, in the affirmative, what they can teach concerning the 
tighter merging occurring in such a situation.

Noncommutative field theories have been widely explored in recent years.
Although their basic motivation relies, in our opinion, by their relation
with string theories \cite{cds,dh,seiwi}, they often exhibit curious 
new features and are therefore
fascinating on their own \cite{minwa,suski}.

The simplest way of turning ordinary theories into non-commutative ones is 
to replace  the 
usual multiplication of fields in the Lagrangian with the Moyal
$\star$-product. 
This product is constructed by means of a real antisymmetric matrix 
$\theta^{\mu\nu}$ which parameterizes  non-commutativity of Minkowski 
space-time:
\beq
\label{alge}
[x^\mu,x^\nu]=i\theta^{\mu\nu}\quad\quad\quad\quad \mu,\nu=0,..,D-1.
\eeq
The $\star$-product of two fields $\phi_1(x)$ and $\phi_2(x)$ can be defined
by means of Weyl symbols
\beq
\label{star}
\phi_1\star\phi_2(x)=\int \frac{d^Dp\, d^Dq}{(2\pi)^{2D}}\exp \lq -\frac{i}2 \, 
p_{\mu}\theta^{\mu\nu}q_{\nu}\rq \exp(ipx) \tilde \phi_1(p-q) \tilde \phi_2(q).
\eeq
The resulting action
makes obviously the theory non-local. 
       
\smallskip
 
A particularly interesting situation occurs in $U(N)$ gauge theories 
defined in 
one-space, one-time dimensions ($YM_{1+1}$).

The classical Minkowski action reads 
\beq 
\label{action}
S=-\ \frac1{4} \int d^2x\,Tr \Big( F_{\m\n} \star  F^{\m\n} \Big)
\eeq
where the field strength $F_{\m\n}$ is given by
\beq
F_{\m\n}=\p_\m A_\n -\p_\n A_\m -ig (A_\m\star A_\n - A_\n\star A_\m)
\eeq
and $A_\m$ is a $N\times N$ hermitian matrix.

The action in Eq.~\re{action} is invariant under $U(N)$
non-commutative gauge transformations 
\beq
\label{gauge}
\de_\l A_\m= \p_\m \l -ig (A_\m\star\l -\l\star A_\m) \,.
\eeq
We quantize the theory in the light-cone gauge $n^{\mu}A_{\mu}\equiv A_{-}=0$,
the vector $n_{\mu}$ being light-like, $n^{\mu}\equiv\frac{1}{\sqrt 2}(1,-1).$ 
This gauge is particularly convenient since
Faddeev-Popov ghosts decouple even in a non-commutative context \cite{sheikh},
while the field tensor is linear in the field with only one non vanishing
component $F_{-+}=\p_{-} A_{+}.$

In this gauge two
{\it different} 
prescriptions are obtained for the vector propagator in momentum space, namely
\beq
\label{hooft}
D_{++}=i\ [k_{-}^{-2}]_{PV}
\eeq
and
\beq
\label{mand}
D_{++}=i\ [k_{-}+i\epsilon k_{+}]^{-2},
\eeq
$PV$ denoting the Cauchy principal value.
The two expressions above are usually referred in the literature as 't Hooft \cite{hoo}
and Wu-Mandelstam-Leibbrandt ($WML$) \cite{wu,wum} propagators. They correspond to 
two different ways of quantizing the theory, namely by means of a light-front
or of an equal-time algebra \cite{bbg,bdbg} respectively and, obviously, coincide
with the ones in the commutative case.

The $WML$ propagator can be Wick-rotated, thereby
allowing for an Euclidean treatment. A smooth continuation of the propagator to the
Euclidean region
is instead impossible when using the $PV$ prescription.

\smallskip
In the commutative case, a perturbative calculation for a closed Wilson loop, 
computed with the 't Hooft's propagator, coincides 
with the exact expression obtained 
on the basis of a purely geometrical procedure \cite{boul,daul}
\beq
\label{looft}
{\cal W}= \exp (-\frac{1}{2}g^2 N {\cal A}).
\eeq
The use instead of the $WML$ propagator leads to a different, genuinely perturbative
expression in which topological effects are disregarded \cite{stau,anlu}
\beq
\label{wml}
{\cal W}_{WML}= \frac{1}{N}
\exp (-\frac{1}{2}g^2 {\cal A}) L^{(1)}_{N-1}( g^2 {\cal A}),  
\eeq
$L^{(1)}_{N-1}$ being a Laguerre polynomial.

One can inquire to what extent these considerations can be generalized to a
non-commutative $U(N)$ gauge theory, always remaining in 1+1 dimensions.
This was explored in ref.\cite{bnt} by performing a fourth order perturbative calculation
of a closed Wilson loop.

In the non-commutative case the  Wilson loop can be defined by means of the
Moyal product as \cite{iikk,gross,alvar}  
\beq
\label{wloop}
\WW[C]=\frac{1}{N}\int \DD A \, e^{iS[A]} \int d^2x\,\tr P_{\star} \exp \lp
ig \int_C A_{+} 
(x+\xi(s))\, d\xi^{+}(s)\rp \,,
\eeq
where $C$ is a closed contour in non-commutative space-time
parameterized by $\xi(s)$, with $0 \leq s \leq 1$, $\xi(0)=\xi(1)$ and $P_\star$
denotes non-commutative path ordering along $x(s)$ from left to right
with respect to increasing $s$ of $\star$-products of functions.
Gauge invariance requires integration over coordinates, which is
trivially realized when considering  vacuum averages \cite{dorn}.
 
The perturbative expansion of $\WW [C]$,
expressed by Eq.~\re{wloop}, reads 
\beeq \label{loopert}
\WW   [C]&=&\frac{1}{N}
\sum_{n=0}^\infty (ig)^n \int_{0}^1 ds_1 \ldots
\int_{s_{n-1}}^{1} ds_{n} 
\, \dot{x}_{-}(s_1)\ldots\dot{x}_{-}(s_{n})\no\\
&&\ph{
\sum_{n=0}^\infty (ig)^n } 
\langle 0\left|\tr{\cal T}\lq  A_{+}(x(s_1))
\star\ldots\star    A_{+}(x(s_{n}))\rq \right|0\rangle, 
\eeeq
and it is shown to be an even power series in $g$, so that we can write
\beq
\label{wpert}
\WW   [C]= 1 +g^2 \WW_2+g^4\WW_4+g^6\WW_6+\cdots\, .
\eeq
If we consider $n$ windings around the loop, the result can be easily obtained
by extending the interval $0\leq s \leq n$, $\xi(s)$ becoming a periodic function of $s.$

The main conclusion of \cite{bnt} was that a perturbative Euclidean calculation 
with the WML
prescription is feasible and leads to a regular result.
We found indeed pure area dependence (we recall that invariance under area preserving 
diffeomorphisms holds also in a non-commutative context) 
and continuity in the limit of vanishing non-commutative
parameter. The limiting case of a large non-commutative parameter 
(maximal non-commutativity) is far from trivial: as a matter of fact
the contribution from the non-planar
graph does not vanish in the large-$\theta$ limit at odds with the result
in higher dimensions \cite{minwa}.

More dramatic is the situation when considering the  't Hooft's form of
the free propagator.
In the non-commutative case the presence of the Moyal phase produces singularities
which cannot be cured \cite{bnt}. 
As a consequence 't Hooft's context will not be further considered.

Another remarkable difference between 't Hooft's and WML formulations
in commutative Yang-Mills theories was noticed in \cite{BGV}.
When considering $n$ windings around the closed loop, a non trivial holonomy
concerning the base manifold and  
the fiber ($U(N)$) (Eq.(\ref{scaling}))
took place in the exact solution.
The behaviour of the WML solution was instead fairly trivial (${\cal A}
\to n^2 {\cal A})$, as expected in
a genuinely perturbative treatment. However it is amazing to notice that the
expression (\ref{wml}) with $n$ windings, when restricted to {\it planar} diagrams
becomes
\beq
\label{wuplanar}
\WW^{(pl)}_{WML}=\sum_{m=0}^{\infty}\frac{(-g^2{\cal A}n^2N)^m}{m!(m+1)!}=
\frac{1}{\sqrt{g^2{\cal A}n^2N}}J_1(2\sqrt{g^2{\cal A}n^2N}).
\eeq
Scaling (\ref{scaling}) is recovered.

In the non-commutative case this issue acquires a much deeper interest
thanks to the merging of space-time and ``internal'' symmetries in a large gauge
group $U(\infty)$, or, better, in its largest completion $U_{cpt}({\cal H})$ \cite{har}. 
Also for the WML formulation we expect a non trivial intertwining between $n$ and
$N$, which might help in clarifying some features of this merging.
Actually this is the main motivation of the present research.

Lacking a complete solution, we limit ourselves to a perturbative context. 
A little thought is enough to be
convinced that the function
$\WW_2$ in Eq.~\re{wpert} is reproduced by the single-exchange diagram, 
which is exactly the same as in the commutative $U(N)$ theory.
Actually all planar graphs contributions coincide with the corresponding ones
of the commutative case \cite{stau}, 
being independent of $\theta$ (see Eq.(\ref{wuplanar})). Although they dominate
for large $N$ and $n$, they are a kind of
``constant'' background, which is uninteresting in this context.
Therefore in the following we will concentrate ourselves in 
calculating and discussing the properties of non-planar graphs
$\WW^{(cr)}$ in the $WML$ (Euclidean) formulation.

The contributions $\WW_4^{(cr)}$ and $\WW_6^{(cr)}$ with $n$ windings
will be presented in detail.
At $\theta=0$ the commutative result is recovered, together with its
trivial perturbative scaling, the result being
continuous (but probably not analytic there).

Surprisingly, at $\theta=\infty$ and at ${\cal O}(g^4)$, we recover  
the non trivial scaling
law (\ref{scaling}) of the exact solution in the commutative case;
however, for the sake of clarity, we stress that such a scaling 
is here realized in a quite
different mathematical expression.
At ${\cal O}(g^6)$ the scaling receives corrections,
decreasing at large $n$; as a consequence we can say that
it holds only at
large $\theta$ and large $n$. We also realize that diagrams with a single
crossing of propagators dominate, making possible the extension to higher perturbative
orders. This evidence is partly based on a numerical evaluation of an integral
occurring in the calculation of diagrams with a double crossing (see Appendix B).

We present arguments in favour of the persistence of such a scaling in
the limits $(n,N,\theta) \to \infty$ at any perturbative order and eventually
succeed in summing the related perturbative series.

As soon as we move
away from the extreme values $\theta=0,\theta=\infty$, 
corrections appear which are likely to
interpolate smoothly between small-$\theta$ and large-$\theta$ behaviours. 

\smallskip

In Sect.2 we present the ${\cal O}(g^4)$ calculation; the ${\cal O}(g^6)$
results are reported in Sect.3 together with our conjecture concerning the
leading terms at large $n,N$ and $\theta$ at any perturbative order.
The details of the calculations are deferred to the Appendices.
Final considerations are 
discussed in the Conclusions.

\section{The fourth order calculation}

We concentrate 
our attention on $\WW_4^{(cr)}$ and resort to an Euclidean
formulation, generalizing to $n$ windings the results reported in \cite{bnt}.

By exploiting the invariance of $\WW$ under area-preserving diffeomorphisms,
which holds also in this non-commutative context, we consider the simple 
choice of a circular contour 
\beq
\label{cerchio}
x(s)\equiv(x_1(s),x_2(s))=r(\cos(2\pi s),\sin(2\pi s)).
\eeq

Were it not for the presence of the Moyal phase, a tremendous simplification
would occur between the factor in the measure $\dot{x}_{-}(s)\dot{x}_{-}(s')$
and the basic correlator $<A_{+}(s)A_{+}(s')>$ \cite{stau}. The Moyal phase can be
handled in an easier way if we perform a Fourier transform, namely if
we work in the momentum space. The momenta are chosen to be Euclidean
and the non-commutative parameter imaginary $\theta\to i \theta$. 
In this way all the phase factors do not change their character.

We use WML propagators in the Euclidean form $(k_1-ik_2)^{-2}$ and parameterize
the vectors introducing polar variables in order to perform symmetric integrations
\cite{wu,stau}. Then we are led to the expression
\begin{eqnarray}  
\label{crociato}
\WW_4^{(cr)} &=& \, r^4 \, \int_0^{n} d{s_1} \int_{s_1}^{n} d{s_2} \int_{s_2}^{n} 
d{s_3 }  \int_{s_3}^{n} d{s_4} \times \nonumber \\
&&\int_0^{\infty } \, {{dp}\over {p}} {{dq}\over {q}} 
\, \int_0^{2 \pi } d\psi \, d\chi \, \exp({- 2 i (\psi + \chi )}) 
\exp({2 i p \sin \psi \sin \pi (s_1 - s_3 )})\times \nonumber  \\
&& \exp({2 i q \sin \chi \sin \pi (s_2 - s_4 )})  \exp({i {{\theta }\over {r^2}} 
p \ q\ \sin [\psi - \chi + \pi (s_2 + s_4 -  s_1 - s_3)] })\nonumber \\
&=& {\cal A}^2 \ F(\frac{\theta}{{\cal A}},n).
\end{eqnarray}

Integrating over $\psi$ and $p$, we get, after a trivial rescaling
\begin{equation}  
\label{integrato}
\WW_4^{(cr)} \, = \, \pi r^4 n^4 \int [ds]_4 \, 
\int_0^{\infty } {{dq} \over {q}} \, 
\ointop_{|z|=1}\, {{dz} \over {i {{z}^3}}} \, \, 
e^{-\ q  \sin [n \pi (s_4 - s_2 )] (z - {{1} \over {z}})} \, 
{{1 - {{\gamma } \over {z}} e^{- i n \pi \sigma }} 
\over {1 - \gamma z e^{i n \pi \sigma }}},
\end{equation}
where  $\sigma = s_1 + s_3 - s_2 - s_4 $ and
$$ \gamma = \frac {\theta q }{2 r^2 \sin [n \pi (s_3 - s_1 )]},\qquad\qquad
\int [ds]_4 = \int_0^1 ds_1 \int_{s_1 }^1 ds_2 \int_{s_2 }^1 ds_3 \int_{s_3 }^1 ds_4 .$$ 

We can further integrate over $z$, obtaining series of Bessel functions.
Integration over $q$ and resummation of the series \cite{prudni} lead to
\begin{eqnarray}  
\label{reintegrato}
\WW_4^{(cr)} \,
&=&2n^4{\cal A}^2 \int [ds]_4\left[\frac{1}{2}+\frac{2}{\beta^2}
\left(\exp[i\beta \sin\alpha]-1-i\beta \sin \alpha\right)\right]\\ \nonumber
  &=&\, 2n^4{\cal A}^2 \int [ds]_4
\left[\frac{1}{2}+\frac{2}{\beta^2}\sum_{m=2}^{\infty}\frac{(i\beta \sin \alpha)^m}
{m!}\right],
\end{eqnarray}
where
\begin{equation}
\label{defini}
\alpha=n\pi (s_1+s_3-s_2-s_4),\qquad \beta=\frac{4{\cal A}}{\pi \theta}
\sin[n\pi (s_4-s_2)]\sin[n\pi(s_3-s_1)].
\end{equation}

It is an easy calculation to check that the function $F$ is continuous (but
probably not analytic) at 
$\theta=0$ with $F(0)=\frac{n^4}{24}$, exactly corresponding to the value
of the commutative case obtained with the WML propagator \cite{stau}.

The first order correction in $\theta$ can also be singled out
\begin{equation}  
\label{piccolo}
\WW_4^{(cr)} \, \simeq 2n^4{\cal A}^2 \int [ds]_4
\left[\frac{1}{2}-\frac{2i}{\beta}\sin\alpha\right].
\end{equation}
The calculation is sketched in Appendix A and the result is
\begin{equation}
\label{respic}
\WW_4^{(cr)}\simeq \frac{n^4{\cal A}^2}{24}+
i \theta \frac{n^3{\cal A}}{4}.
\end{equation}
One might recover the trivial scaling ${\cal A}\to {\cal A}n^2$ provided
$\theta \to \theta n$; 
however this is ruled out by the large-$\theta$
behaviour we are going to explore.

The large-$\theta$ behaviour can be obtained starting from Eq.(\ref{reintegrato});
the first terms in the expansion turn out to be
\begin{equation}
\label{granteta}
\WW_4^{(cr)} \, =-\frac{n^2{\cal A}^2}{8\pi^2}+i\frac{n^3{\cal A}^3}{8\pi^2\theta}+
\frac{8n^4{\cal A}^4}{3\pi^2\theta^2}\left(\frac{1}{256}+\frac{175}{3072}\frac{1}
{n^2\pi^2}\right)+{\cal O}(\theta^{-3}).
\end {equation}

We notice that the large-$\theta$ limit (first term in Eq.(\ref{granteta})) 
obeys the scaling (\ref{scaling}),
which, in the commutative case, was present in the exact solution
for the gauge group $U(N)$. This scaling is different from the trivial one at
$\theta=0$. 

\section{The sixth order calculation and beyond}

The motivation for exploring the sixth order is to see whether the scaling
law we have found in the fourth order result
at $\theta=\infty$ still persists in higher orders. In the affirmative case
one would be strongly encouraged to resum the series in order to inquire
about the persistence of such a scaling beyond a perturbation expansion. This, in turn,
might have far-reaching consequences on the interpretation of the theory
in the extreme non-commutative limit.

We organize the sixth order loop calculation according to the possible
topologically different diagrams one can draw. If we order the
six vertices on the circle from 1 to 6, we denote by $\WW_{(ij)(kl)(mn)}$
the contribution of the graph corresponding to three propagators joining
the vertices $(ij),(kl),(mn)$, respectively. Thus  $\WW_{(14)(25)(36)}$
corresponds to the maximally crossed diagram 
(i.e. the one in which all propagators cross);
then we have three diagrams with double crossing, namely $\WW_{(14)(26)(35)},
\WW_{(13)(25)(46)}$ and $\WW_{(15)(24)(36)}$. Finally we have six diagrams
with a single crossing $\WW_{(12)(35)(46)}, \WW_{(16)(24)(35)}, \WW_{(15)(23)(46)},
\WW_{(15)(26)(34)}, \WW_{(13)(26)(45)}$ and $\WW_{(13)(24)(56)}$.
Diagrams without any crossing are not interesting since they are not affected
by the Moyal phase; they indeed coincide with the corresponding ones in the
commutative case.

The diagrams with a single crossing can be fairly easily evaluated; surprisingly
the most difficult diagrams are the ones with double crossing. All integrations
can be performed analytically, but a single one concerning the doubly crossed
diagrams, which has been performed numerically.

The details of such a heavy calculation are described in Appendix B.
Here we only report the starting point and the final results.

\noindent
As an example of singly crossed diagram we consider $\WW_{(16)(24)(35)}$
\begin{eqnarray}  
\label{scrossed}
\WW_{(16)(24)(35)}&=& \,- r^6 N n^6\, \int [ds]_6
\int_0^{\infty } \, {{dp}\over {p}} {{dq}\over {q}} {{dk}\over {k}} 
\, \int_0^{2 \pi } d\phi \, d\chi \, d\psi\, \exp({- 2 i (\phi + \chi +\psi )})
\times \\ \nonumber 
&&\exp\left({2 i p \sin \phi \sin n\pi s^-_{16}}
+{2 i q \sin \psi \sin n\pi s^-_{24}}+
{2 i k \sin \chi \sin n\pi s^-_{35} }\right) \times \\ \nonumber 
&&\exp\left(i {{\theta }\over {r^2}} 
q\ k\ \sin [\psi - \chi + n\pi (s^+_{35} -  s^+_{24})] \right),
\end{eqnarray}
where $s^{\pm}_{ij}=s_i\pm s_j$.

\noindent
The doubly crossed diagram $\WW_{(15)(24)(36)}$ leads to the expression
\begin{eqnarray}  
\label{dcrossed}
\WW_{(15)(24)(36)}&=& \,- r^6 N n^6\, \int [ds]_6
\int_0^{\infty } \, {{dp}\over {p}} {{dq}\over {q}} {{dk}\over {k}} 
\, \int_0^{2 \pi } d\phi \, d\chi \, d\psi\, \exp({- 2 i (\phi + \chi +\psi )})
\times \\ \nonumber 
&&\exp({2 i p \sin \phi \sin n\pi s^-_{15} })
\exp({2 i q \sin \psi \sin n\pi s^-_{24}})\times \\ \nonumber
&&\exp(2 i k \sin \chi \sin n\pi s^-_{36})  
\exp\Bigg(i {{\theta }\over {r^2}}\Big( 
p\ k\ \sin [\phi - \chi + n\pi (s^+_{36} -  s^+_{15}) ]+\\ \nonumber
&& q\ k\ \sin [\psi - \chi + n\pi (s^+_{36} -  s^+_{24})] \Big) \Bigg).
\end{eqnarray}

\noindent
Finally the maximally crossed diagram  $\WW_{(14)(25)(36)}$ reads
\begin{eqnarray}  
\label{mcrossed}
\WW_{(14)(25)(36)}&=& \,- r^6 N n^6\, \int [ds]_6
\int_0^{\infty } \, {{dp}\over {p}} {{dq}\over {q}} {{dk}\over {k}} 
\, \int_0^{2 \pi } d\phi \, d\chi \, d\psi\, \exp({- 2 i (\phi + \chi +\psi )})
\times \\ \nonumber 
&&\exp({2 i p \sin \phi \sin n\pi s^-_{14} })
\exp({2 i q \sin \psi \sin n\pi s^-_{25} })\times \\ \nonumber 
&&\exp({2 i k \sin \chi \sin n\pi s^-_{36}}) 
\exp\Bigg(i {{\theta }\over {r^2}}\Big( p\ q\ 
\sin [\phi - \psi + n\pi (s^+_{25} -  s^+_{14})]+ \\ \nonumber
&& p\ k\ \sin [\phi - \chi + n\pi (s^+_{36} -  s^+_{14})]+
q\ k\ \sin [\psi - \chi + n\pi (s^+_{36} -  s^+_{25})] \Big) \Bigg).
\end{eqnarray}

We notice that the $U(N)$ factor is the same in all the three configurations.

The sum of the diagrams with a single crossing and $n$ windings contribute 
at $\theta=\infty$ with
the following expression
\begin{equation}
\label{singlec}
\WW^{(1)}(\theta=\infty)= \frac{{\cal A}^3 N n^4}{24\pi^2}\left(1-
\frac{6}{n^2\pi^2}\right).
\end{equation}

The maximally crossed diagram in turn leads to
\begin{equation}
\label{max}
\WW^{(3)}(\theta=\infty)=\ -\frac{{\cal A}^3 N n^2}{64\pi^4}.
\end{equation}

Finally the diagrams with double crossing give
\begin{equation}
\label{dc}
\WW^{(2)}(\theta=\infty)=\ \frac{{\cal A}^3 N n^2}{12\pi^4}(1+
0.2088).
\end{equation}
As we have anticipated, the last term has been evaluated numerically. 
Its $n$-dependence has been checked up to $n=6$,
within the incertitude due to the
numerical integration (see Appendix B). 

Summing together all the contributions of diagrams with crossed propagators, we get
\begin{equation}
\label{tot}
\WW_6^{(cr)}(\theta=\infty)=\ \frac{{\cal A}^3 N n^4}{24\pi^2}\left(1-\frac{1}{n^2\pi^2}
(\frac{35}{8}-0.4176)\right).
\end{equation}

We remark that the leading term at large $n$ $$\WW_6^{(cr)}(\theta=\infty)
\simeq \frac{{\cal A}^3
N n^4}{24\pi^2}$$ exhibits the scaling
(\ref{scaling}). It comes only from diagrams with a single crossing. Diagrams with
such a topological configuration can also be computed in higher orders; for
instance, at ${\cal O}(g^8)$ they lead to the result
\beq
\label{higher}
\WW_8^{(cr)}(\theta=\infty)\simeq -\frac{{\cal A}^4N^2n^6}{192\pi^2}+{\cal O}(n^4).
\eeq
The integral over the loop variables 
provides a factor $n^{-2}$, turning the trivial $n^8$, due to
the kinematical rescaling, into the factor $n^6$. 
Details are reported in Appendix C.

We are led to argue that the dominant term at the
$(2m+4)$-th perturbative order increases with $n$ not faster than $n^{2m+2}$.
In turn it exhibits the highest $U(N)$ contribution, behaving like $N^m$
\beq
\label{arg}
g^{2m+4}\WW_{2m+4}^{(cr)}(\theta=\infty)
\simeq {\cal K}_m (nN)^m \, (g^2{\cal A}n)^{m+2},
\eeq
which obeys the scaling (\ref{scaling}).

Further we conjecture that diagrams with a single crossing dominate; then the
weights ${\cal K}_m$ can be evaluated (see Appendix D) and lead to
\beq
\label{conject}
g^{2m+4}\WW_{2m+4}^{(cr)}(\theta=\infty)
\simeq -\frac{(g^2{\cal A}n)^2}{4\pi ^2}\,\frac{1}{m!}
\frac{(-g^2{\cal A} N n^2)^m}{(m+2)!};
\eeq
the related perturbative series can be easily resummed
\beq
\label{resu}
\WW^{(cr)}(\theta=\infty)=-\frac{g^2{\cal A}}{4\pi^2N}\, J_2(2\sqrt{g^2{\cal A}
n^2N}).
\eeq
If we compare Eq.(\ref{conject}) with the corresponding term due to planar diagrams,
which are insensitive to $\theta$  
(see Eq.(\ref{wuplanar})), we notice that, in the 't Hooft's limit $N\to \infty$ with
fixed $g^2N$, the planar diagrams dominate by a factor $n^2 N^2$, as expected.

Our conjecture is open to more thorough perturbative tests as well as to possible 
non-perturbative derivations which might throw further light on its ultimate
meaning and related consequences. For recent papers on non-perturbative approaches,
see \cite{luca,bie,sza}.

\section{Conclusions}

\noindent
Summarizing our perturbative investigation, we can say that, when winding
$n$-times around the Wilson loop the non-Abelian nature of the gauge group
in the non-commutative
case is felt, even in a perturbative calculation making use of the $WML$
prescription for the vector propagator. This is due to the merging of 
space-time properties with
``internal'' symmetries in a large invariance group $U_{cpt}({\cal H})$ \cite{har,sza}.

One gets the clear impression that in a non-commutative formulation
what is really relevant are not separately the space-time properties
of the ``base'' manifold and of the ``fiber'' $U(N)$, but rather
the overall algebraic structure of the resulting invariance group
$U_{cpt}({\cal H})$. To properly understand its topological features is
certainly beyond any perturbative approach. Rather one should possibly resort
to suitable ${\cal N}$-truncations of the Hilbert space in the form 
of matrix models leading to the invariance groups $U({\cal N})$.

It is not clear how many perturbative features might eventually be
singled out in those contexts, especially in view of the difficulty in performing the
inductive limit ${\cal N}\to \infty$. 

For this reason we think
that our perturbative results are challenging.
They indicate that the intertwining between $n$, controlling the
space-time geometry, and $N$, related to the gauge group, is far
from trivial. The presence of corrections to the scaling laws
occurring at $\theta=0$ and at $\theta=\infty$, while frustrating
at a first sight in view of a generalization to all values
of $\theta$, might be taken
instead as a serious indication
that $n$ and $N$ separately are not perhaps the best parameters to be chosen
unless large values for both (and for $\theta$!) are considered. In such a
situation, perhaps surprisingly, the relation (\ref{scaling}) is recovered.

Eqs.(\ref{wuplanar},\ref{conject}) 
are concrete realizations of the more general structure
\beq
\label{gener}
\WW_{2m+4}=({\cal A}n^2 N)^{m+2}\, f_m(n,N),
\eeq
$f_m$ being a symmetric function of its arguments. We stress that Eq.(\ref{wuplanar}) 
concerns only {\it planar} diagrams; crossed graph contributions
in the commutative case cannot be put in the form (\ref{gener}) and violate the relation
(\ref{scaling}) \footnote{The structure (\ref{gener}) is shared
also by the exact geometrical solution of the commutative case \cite{BGV}.}. 
In the non-commutative case, for large $n,N$ and maximal
non-commutativity ($\theta=\infty$), the structure (\ref{gener})
is instead restored for the leading contribution
of {\it crossed} diagrams. The presence of the function $f_m$ in the $WML$ context
might be thought as a sign of the merging of space-time
and internal symmetries. 

All these difficult, but intriguing questions are worthy in our opinion
of thorough investigations and promise further exciting, unexpected developments.

\vskip 1.0truecm

\noindent
$\bf Acknowledgements$

We thank Matteo Viel for help in the numerical calculation. One of us (AB)
acknowledges a partial support by the European Community network HPRN-CT-2000-
00149.

\appendix
\section{Small $\theta$ Limit}
The first significant term in the small $\theta$ expansion is 
the second one 
in  eq.(\ref{piccolo}), that is the ${\cal O}(\theta)$ 
term in $\WW_4^{(cr)}$,
\begin{eqnarray}  
\label{st1}
\WW_\theta&=&-in^4{\cal A}\pi\theta \int_0^1[ds]
{\sin[n \pi (s_1+s_3-s_2-s_4)]\over\sin[n \pi (s_4-s_2)]
\sin[n\pi (s_3-s_1)]}\nonumber\\
&&\equiv -in^4{\cal A}\pi\theta I\ \ ,
\end{eqnarray}
with measure $[ds]=ds_1 ds_2 ds_3 ds_4 \theta(s_4-s_3)\theta(s_3-s_2)
\theta(s_2-s_1)$. Integrating in $ds_1$ and $ds_4$ leads to

\begin{eqnarray}  
\label{st2}
I&=& {1\over n^2 \pi^2}\int_0^1ds_2 ds_3\theta(s_3-s_2)\Bigl\{-n \pi 
\cos[2 n \pi (s_3-s_2)]\times\Bigr.\nonumber\\
&&\Bigl. \Biggl[s_2 \log\left|{\sin n\pi s_2\over \sin n \pi(s_3-s_2)}
\right|+(1-s_3)\log\left|{\sin n\pi s_3\over \sin n \pi(s_3-s_2)}
\right|\Biggr]\Bigr.\nonumber\\
&&\Bigl.+\sin[2 n \pi (s_3-s_2)]\Biggl[-n^2 \pi^2 s_2 (1-s_3)+
\log\left|\sin n \pi (s_3-s_2)\right|\log\left|{\sin n \pi (s_3-s_2)
\over\sin n\pi s_2 \sin n\pi s_3}\right|\Biggr.\Bigr.\nonumber\\
&&\Biggl.\Bigl. +\log |\sin n \pi s_2 |\log |\sin n \pi s_3 |\Biggr]
\Bigr\}\equiv I_1+I_2\ \ ,
\end{eqnarray}
where $I_1$ and $I_2$ refer to the first and second square brackets 
in (\ref{st2}), respectively.
The two integrals in $I_1$ coincide. They can be easily perfomed leading to
\begin{equation}
\label{st3}
I_1=-\left( {1\over 6 n \pi}+{1\over 4 n^3 \pi^3}\right)\ .
\end{equation}
Concerning $I_2$, the first term is trivial, and provides us with a factor
$1/(8 n^3 \pi^3)-1/(12 n\pi)$, whereas in the remaining integrals 
it is more convenient to integrate first on one variable and then to add
the integrands together before performing the final integration, {\it i.e.}
\begin{eqnarray}
\label{st4}
I_2 &=&{1\over 8 n^3 \pi^3}-{1\over 12 n\pi}-{1\over 2 n^3 \pi^3}
\int_0^1 ds\sin n\pi s\times\nonumber\\
&&\Bigl[2 n \pi s\cos n\pi s\log|\sin n \pi s |+\sin n \pi s 
(\log |\sin n \pi s |-1)\Bigr]\nonumber\\
&=&{1\over 4 n^3 \pi^3}-{1\over 12 n\pi}\ .
\end{eqnarray}
Adding (\ref{st3}) and (\ref{st4}) and taking into account 
Eq.(\ref{st1}),
Eq.(\ref{respic}) follows.

\section{Sixth Order Calculation }

\subsection{The singly crossed diagram}  
We show in some detail the formulas for $\WW_{(16)(24)(35)}$, the other five
diagrams being simply obtainable by renaming the variables.

Integrating (\ref{scrossed}) over $\phi$ and $p$ we recover an expression analogous to
~\re{crociato}. Therefore we can use the result obtained at ${\cal O}(g^4)$ to get
\begin{eqnarray}  
\label{magic}
\WW_{(16)(24)(35)} \,
&=& -2 n^6 {\cal A}^3 N \int [ds]_6 \left[\frac{1}{2}+\frac{2}{{\beta'}^2}
\left(\exp[i\beta' \sin\alpha']-1-i\beta' \sin \alpha'\right)\right]
\end{eqnarray}
where now 
\begin{equation}
\label{newdefini}
\alpha'=n\pi (s_2+s_4-s_3-s_5),\qquad \beta'=\frac{4{\cal A}}{\pi \theta}
\sin[n\pi (s_4-s_2)]\sin[n\pi(s_5-s_3)].
\end{equation}
The large-$\theta$ limit is easily derived from this formula; summing all
the singly crossed diagrams we find
\beq
\label{scd}
- \, {\cal A}^3 \, N \, n^6  \, \Big[ {{1}\over {4 \pi^4 n^4 }} \, 
- \, {{1}\over {24 \pi^2 n^2}}\Big]. 
\eeq

\subsection{The doubly crossed diagram}

Integrating Eq.(\ref{dcrossed}) over $\phi$ and $p$, and then over $\psi$ and $q$, we get
\begin{eqnarray}  
\label{duecast}
\WW_{(15)(24)(36)} \, &=& \, - r^6 N n^6 {\pi }^2 \int [ds]_6  \, 
\int_0^{\infty } {{dk} \over {k}} \, 
\ointop_{|z|=1}\, {{dz} \over {i {{z}^3}}} \, \, 
e^{-\ k  \sin [n \pi (s_6 - s_3 )] (z - {{1} \over {z}})} \nonumber \\
&& {{1 - {{\gamma }' \over {z}} e^{- i n \pi \sigma' }} 
\over {1 - \gamma' z e^{i n \pi \sigma' }}} \, \times \, {{1 - {{\gamma }'' \over {z}} e^{- i n \pi \sigma'' }} 
\over {1 - \gamma'' z e^{i n \pi \sigma'' }}}
\end{eqnarray}
where
$$
\sigma' = s_1 + s_5 - s_3 - s_6 \, \qquad \, \gamma' = \frac {\theta k }{2 r^2 \sin [n \pi (s_5 - s_1 )]} 
$$
$$
\sigma'' = s_2 + s_4 - s_3 - s_6 \, \qquad \, \gamma'' = \frac {\theta k }{2 r^2 \sin [n \pi (s_4 - s_2 )]} 
$$
We consider the identity $e^{-\ k  \sin [n \pi (s_6 - s_3 )] (z - {{1} \over {z}})} \, 
\equiv \, [(e^{-\ k  \sin [n \pi (s_6 - s_3 )] (z - {{1} \over {z}})} - 1) + 1]$ 
in ~\re{duecast}; in the first term it is possible to send $\theta$ to infinity in the
integrand, obtaining the result
\begin{eqnarray}   
\label{conj}
- {{r^6 \, N \, n^6 \, {\pi}^3}\over {3}} \, \int [ds]_6 \, e^{2 \pi i (2 s_3 + 2 s_6 - s_1 - s_5 - s_2 - s_4 )}.
\end{eqnarray}
The other contribution can be exactly integrated over $z$ and $k$, leading
to the sum of two expressions 
\begin{eqnarray}
\label{cos3}
- {{r^6 \, N \, n^6 \, 2 \, {\pi}^3}\over {3}} \, \int [ds]_6 \, \exp({- i (\lambda \, + \, \omega )}) \, [\cos({\lambda \, - \, \omega })]^3  
\end{eqnarray}
and
\begin{eqnarray}
\label{malefic}
{{i \, r^6 \, N \, n^6 \, 2 \, {\pi}^3}\over {3}} \, \int [ds]_6 \, \exp({- i (\lambda \, + \, \omega )}) \, {{\Big( 1 + |{{d}\over {c}}| \exp({i({\lambda \, - \, \omega })})\Big)}\over {\Big( 1 - |{{d}\over {c}}| \exp({i({\lambda \, - \, \omega })})\Big)}} \, [\sin({\lambda \, - \, \omega })]^3, 
\end{eqnarray}
where
$$
c \, = \, - \sin [n \pi (s_1 - s_5 )] \, \exp({- i n \pi \sigma' }) \, = \, |c| \, \exp({i \omega }),
$$
$$
d \, = \, - \sin [n \pi (s_2 - s_4 )] \, \exp({- i n \pi \sigma'' }) \, = \, |d| \, \exp({i \lambda }).
$$
The integrals ~\re{conj} and ~\re{cos3} can be easily computed;
when summed with the corresponding ones from $\WW_{(14)(26)(35)}$ and $\WW_{(13)(25)(46)}$, they give 
\beq
\label{dcr}
{{{\cal A}^3 \, N \, n^2 }\over {12 \, {\pi }^4 }}.
\eeq

Expression ~\re{malefic} instead, together with the corresponding ones 
from $\WW_{(14)(26)(35)}$ and $\WW_{(13)(25)(46)}$, is difficult to deal with. 
We can prove their sum is real and have evaluated such a sum numerically, 
for $n = 1,..., 6$. We present the result in the form
\beq
\label{numer}
{{4 \, \pi \, r^6 \, N \, n^4}\over {3}} \, \times \, J_{NUM},
\eeq
where

$$
J_{NUM} (n = 1) \, = \,  1.32236(80 \pm 37) \times 10^{- 3}   
$$
$$
J_{NUM} (n = 2) \, = \,  0.330(49 \pm 16) \times 10^{- 3} \qquad {{J_{NUM} (n = 1)}\over {4}} \, = \, 0.330592(01 \pm 93) \times 10^{- 3}      
$$
$$
J_{NUM} (n = 3) \, = \,  0.146(97 \pm 35) \times 10^{- 3} \qquad {{J_{NUM} (n = 1)}\over {9}} \, = \, 0.146929(80 \pm 41) \times 10^{- 3}      
$$
$$
J_{NUM} (n = 4) \, = \,  0.08(17 \pm 29) \times 10^{- 3} \qquad {{J_{NUM} (n = 1)}\over {16}} \, = \, 0.082648(00 \pm 23) \times 10^{- 3}      
$$
$$
J_{NUM} (n = 5) \, = \,  0.05(10 \pm 40) \times 10^{- 3} \qquad {{J_{NUM} (n = 1)}\over {25}} \, = \, 0.052894(72 \pm 15) \times 10^{- 3}      
$$
$$
J_{NUM} (n = 6) \, = \,  0.03(88 \pm 79) \times 10^{- 3} \qquad {{J_{NUM} (n = 1)}\over {36}} \, = \, 0.036732(45 \pm 10) \times 10^{- 3}      
$$

All the errors are three standard deviations. 
Within the numerical error, $J_{NUM}$ scales as ${{1} / {n^2}}$.

\subsection{The maximally crossed diagram}

Integrating Eq.(\ref{mcrossed}) over $\phi$ and $p$, and then over $\chi$ and $k$, 
we get, after a simple rescaling
\begin{eqnarray} 
\label{storic}
\WW_{(14)(25)(36)} &=& \, - r^6 \, N \, n^6 \, 2 {\pi}^2 \, \int [ds]_6 \,\int_0^{\infty } \,  {{dq}\over {q}} \, \int_0^{2 \pi } \, d\psi \, \exp({- 2 i \psi })\times \nonumber  \\
&& \exp({4 i {{q r^2}\over {\theta}} { \sin \psi \sin [n \pi (s_2 - s_5 )]}}) \, \Bigg[ \, \, {{1}\over {2}} \, {{\bar{\alpha}}\over {\alpha}} \,{{\bar{\beta}}\over {\beta}} \, + \, \nonumber \\
&& {{{\theta}^2}\over {8 r^4 {\alpha}^2 {\beta}^2}} \left( \exp\Big({{{4 i r^2 }\over {\theta}} \, {\rm Im} (e^{i n \pi {\sigma}'''} \bar{\alpha } \beta)}\Big) \, - \, {{4 i r^2 }\over {\theta}} \, {\rm Im} (e^{i n \pi {\sigma}'''} \bar{\alpha } \beta) \, - 1 \, \right)  \Bigg],
\end{eqnarray}     
where ${\sigma}''' = s_1 + s_4 - s_3 - s_6$, the bars denote complex conjugation
and
$$
\alpha \, = \, \sin [n \pi (s_1 - s_4 )] \, + \, q \, \exp\ i({\psi + n \pi (s_1 + s_4 - s_2 - s_5 )}), 
$$
$$
\beta \, = \, \sin [n \pi (s_3 - s_6 )] \, - \, q \, \exp\ i({\psi + n \pi (s_3 + s_6 - s_2 - s_5 )}) . 
$$
We can recognize in (\ref{storic}) the same structure we have found in (\ref{reintegrato}).
We rewrite it as follows:
\begin{eqnarray} 
\label{drago}
&&\WW_{(14)(25)(36)} = \, - r^6 \, N \, n^6 \, 2 {\pi}^2 \, \int [ds]_6 \,\int_0^{\infty } \,  {{dq}\over {q}} \, \int_0^{2 \pi } \, d\psi \, \exp({- 2 i \psi })\times \nonumber  \\
&& \exp({4 i {{q r^2}\over {\theta}} { \sin \psi \sin [n \pi (s_2 - s_5 )]}}) \, \exp({- 2 i ({\gamma }_{\alpha } + \, {\gamma }_{\beta })}) \, \Bigg[ \, {{1}\over {2}} \, \cos({2 n \pi {\sigma }''' \, - \, 2 {\gamma }_{\alpha } \, + \, 2 {\gamma }_{\beta } }) \, + \nonumber \\
&& {{1}\over {\pi i}} \, \int_{\mu - i \infty }^{\mu + i \infty } \, ds \, \Gamma (- s) \, e^{- i {{\pi }\over {2}} s} \, {\Bigg[ {{4 |\alpha | |\beta | r^2 }\over {\theta }} \Bigg]}^{s - 2} \, {\big[ \sin({n \pi {\sigma }''' \, - \, {\gamma }_{\alpha } \, + \, {\gamma }_{\beta } }) \big]}^s \Bigg],\qquad 2<\mu<3,
\end{eqnarray}        
where we have defined $\alpha = |\alpha | \exp({i {\gamma }_{\alpha }})$ and $\beta = |\beta | \exp({i {\gamma }_{\beta }})$. 

One can prove that, in the large-$\theta$ limit, the last integral goes to zero. Then, in this limit, equation (\ref{drago}) can be easily evaluated:
\begin{eqnarray} 
\label{symm}
\WW_{(14)(25)(36)} &{\longrightarrow}_{\theta \to \infty}& \, - {{r^6 \, N \, n^6 \, {\pi}^3}\over {3}} \, \int [ds]_6 \, \Big( e^{ 2 i n \pi (2 s_2 + 2 s_5 - s_1 - s_4 - s_3 - s_6 )} \, + \, \nonumber \\
&& e^{ 2 i n \pi (2 s_1 + 2 s_4 - s_2 - s_5 - s_3 - s_6 )} \, + \, e^{ 2 i n \pi (2 s_3 + 2 s_6 - s_1 - s_4 - s_2 - s_5 )} \Big) \nonumber \\     && \, = \, - \, {{{\cal A}^3 N n^2 }\over {64 {\pi }^4 }}.
\end{eqnarray}
Here we notice that the integrand is completely symmetric in the three propagators 
(14)(25)(36), as it should.   

\section{Higher orders}

First we prove that singly crossed diagrams behave in the large-$\theta$ limit
at least as 
${{1} \over {n^2}}$ or subleading in the limit of a large number of windings $n$. 
We start by realizing that we can always express the integral of 
a generic diagram with $m$ 
propagators and a single crossing generalizing Eq.(\ref{reintegrato})
\beq
\label{ap1}
{\cal I}\equiv \int_{0}^{1}  dt \, \int_{0}^{t}  dz \, \int_{0}^{z}  dy \, \int_{0}^{y}  dx \, \int  [ds]_{2m - 4} \, \cos [2 \pi n (x + z - y - t)],
\eeq
$[ds]_{2m-4}$ being a measure depending on $x,y,z,t$ only through the extremes 
of integration.
As a matter of fact, it is always possible to single out the variables linked to
the propagators which cross, suitably rearranging the other kinematical integrations.
These integrations lead to polynomials
\beq
\label{ap2}
{\cal I}=\int_{0}^{1}  dt \, \int_{0}^{t}  dz \, \int_{0}^{z}  dy \, \int_{0}^{y}  dx \, \sum_{k_1 k_2 k_3 k_4} \, c_{k_1 k_2 k_3 k_4} \, x^{k_1} y^{k_2} z^{k_3} t^{k_4} \, \cos [2 \pi n (x + z - y - t)].
\eeq
Now we perform the change of variables $\alpha = y + x$, $\beta = y - x$, $\gamma= t +z$, $\delta = t - z$
\begin{eqnarray}
\label{ap3}
{\cal I}&=&\int_0^1 d\delta \, \int_{\delta}^{2 - \delta} d\gamma \, 
\int_0^{{{\gamma - \delta }\over {2}}} d\beta \, 
\int_{\beta}^{\gamma - \delta - \beta } d\alpha \\ \nonumber
&&\sum_{q_1 q_2 q_3 q_4} \, {c'}_{q_1 q_2 q_3 q_4} \, {\alpha}^{q_1} {\beta}^{q_2} {\gamma}^{q_3} {\delta}^{q_4} \, \cos [2 \pi n (\beta + \delta )]
\end{eqnarray}
and then integrate over $\alpha$. Changing again variables to $\psi = \beta + \delta$, 
$\xi = \delta - \beta$, we end up with
\beq
\label{ap4}
{\cal I}=\int_0^1 d\psi \, \int_{-\psi}^{\psi} d\xi \int_{\frac{3\psi-\xi}{2}}
^{2-\frac{\psi+\xi}{2}} d\gamma  \, \sum_{p_1 p_2 p_3} \, {C}_{p_1 p_2 p_3} \, {\psi}^{p_1} {\xi}^{p_2} {\gamma}^{p_3} \, \cos [2 \pi n \psi]. 
\eeq     

The integrals over $\xi$ and $\gamma$ can be easily performed
giving, of course, a polynomial in $\psi$
\beq
\label{ap5}
{\cal I}=\sum_r \, {C'}_r \, \int_0^1 \, {\psi}^r \, \cos [2 \pi n \psi]\, d\psi.        \eeq

Integrating by parts, we realize that only even inverse powers of $n$
are produced, starting from $n^{-2}$.

\smallskip

Now we turn our attention to the $U(N)$ factors. A direct computation of the traces
involved in the diagrams with a single, a double or the triple crossing 
(${\cal O}(g^6)$), shows that they all share the common factor $N^2$ (our
normalization being 
$t^0={\bf 1}/{\sqrt N},\ Tr (t^a t^b)=\delta^{ab}; \quad a,b=1,...,N^2-1).$
As the Wilson loop is normalized with $N^{-1}$, at ${\cal O}(g^6)$ the single
factor $N$ ensues.

It is now trivial to realize that any insertion of $m-3$ lines no matter where in
such diagrams provided that further crossings are avoided, produces the factor
$N^{m-3}$. 

\section{Computation of the weights}

In the previous appendix we have shown that the $n$-dependence of singly 
crossed diagrams in the large-$\theta$ limit takes the form 
$\sum_{p=1}^{P} c_p \, n^{- 2 p}$. To find the leading contribution at large $n$ 
we have to evaluate $c_1$. This can be done as follows: at ${\cal O}(g^{2 m + 4})$ 
we start drawing a cross and then add the remaining $m$ propagators in such a way 
they do not further cross. From Eqs.(\ref{ap2}-\ref{ap5}) one can realize that $c_1$ 
is different from zero only for a particular subset of these diagrams: if we label
the four sectors in which the cross divides the circular loop as North (the sector
containing the origin of the loop variables $s_i$), West, South and East, 
then only diagrams with $r$ propagator in the southern sector and $m - r$ in the 
northern one contribute to $c_1$; moreover these contributions are all equal.
Therefore we can evaluate this integral once and then multiply it by the number of
configurations in this subset.                  
 
We choose as representative the diagram with all the $m$ non-intersecting propagators 
in the northern sector, starting from the origin and connecting $s_1$ with $s_2$,
..., $s_{2m-1}$ with $s_{2m}$. In this way the crossed variables are $s_{2m+1}$,
..., $s_{2m+4}$. We obtain the integral
\beq
\label{D1}
{\cal I} = {(- \pi )}^{m + 2} {(g r )}^{2 m + 4}  N^{m} n^{2 m + 4}  
\int_{0}^{1}  dt \, \int_{0}^{t}  dz \, \int_{0}^{z}  dy \, \int_{0}^{y}  dx \,
{{x^{2 m}}\over {(2 m)!}} \cos [2 \pi n (x + z - y - t)].
\eeq

Following the procedure described in appendix C we get
\beq
\label{D2}
{\cal I} = {{{(- \pi )}^{m + 2} {(g r )}^{2 m + 4}}\over {{(2 m)!}}}  N^{m}
n^{2 m + 4}   \int_{0}^{1}  d\psi {{1}\over {(2 m + 1) (2 m + 2)}} 
\psi {(1 - \psi )}^{2 m + 2} \cos [2 \pi n \psi]
\eeq

and finally

\beq
\label{D3}
{\cal I} = - N^{m} {{{(- g^2 {\cal A} \, n^2 ) }^{m + 2}}\over {(2 m + 2)!}}  
\Bigg( {{1}\over {4 {\pi}^2 n^2}} + {\cal{O}}({{1}\over {n^4}})\Bigg) 
\eeq

Now we have to count. We denote by $S_{2r}$ the ways in which the $r$ propagators 
in the southern sector can be arranged without crossing. A little thought provides
the recursive relation
\beq
\label{D4}
S_0=1, S_{2r}=\sum_{k=1}^{r} S_{2k-2} S_{2r-2k},
\eeq
which can easily be solved
\beq
\label{D5}
S_{2r} = {{{2^{2r}} \Gamma (r + {1\over 2})}\over {{\Gamma ({1\over 2})} 
\Gamma (r + 2)}}.
\eeq

The $m - r$ propagators in the northern sector lead to the weight $S_{2(m - r)}$ 
times the number of possible insertions of the origin, namely $[2(m - r) + 1]$. 
The number of relevant diagrams is therefore
\beq
\label{D6}
{\cal N}_m = \sum_{r=0}^m S_{2r}\, S_{2(m - r)} \, [2(m - r) + 1] \, =
 \, {{{2^{2 m + 2}} (m + 1) \Gamma (m + {3\over 2}) }\over {\Gamma ({1\over 2}) 
\Gamma (m + 3)}}.
\eeq  

Multiplying Eqs.(\ref{D3}) and (\ref{D6}) we are led to Eq.(\ref{conject}).

\end{document}